\documentstyle[12pt]{article}
\begin{document}

\title{Parabose Squeezed Operator and Its Applications \thanks{
This project supported by National Natural Science Foundation of China under
Grant 19771077 and LWTZ 1298.}}
\author{{Weimin Yang and Sicong Jing } \\
{\small Department of Modern Physics, University of Science and}\\
{\small Technology of China, Hefei, Anhui 230026, P.R.China}}
\maketitle

\begin{abstract}
By virtue of the parabose squeezed operator, propagator of a parabose
parametric amplifier, explicit form of parabose squeezed number states and
normalization factors of excitation states on a parabose squeezed vacuum
atate are calculated in this letter, which generalize the relevant results
from ordinary Bose statistics to parabose case.
\end{abstract}

\bigskip 

PACS numbers: 03.65.Fd

Key words: parastatistics, squeezed operator, deformed special functions

\section{Introduction}

A fundamental unsolved question in physics is whether all particles in the
nature are necessarily either bosons or fermions. Theoretical inverstigations
of other possibilities in local, relativistic quantum field theory show that
may exist more general particle statistics. Parastatistics was introduced by
Green as an exotic possibility extending the Bose and Fermi statistics
\cite{s1} and for the long period of time the interest to it was rather
academic. Nowadays it finds some applications in the physics of the quantum
Hall effect \cite{s2} and (probably) it is relevant to high temperature
superconductivity \cite{s3}, so more and more attentions are paid to it in
recent years. Even though there are no observed paraparticles in nature, the
possibility exists for unobserved particles which obey the parastatistics.
Some developments in interacting many-particle systems have also shown that
the quasiparticles in such systems may exhibit features far more exotic than
those permitted to ordinary particles, and it appeara quite possible that
parastatistics may be realized in condensed matter physics. In the case of
parastatistics, there is the additional motivation of the possible production
of $p>1$ paraparticles ($p$ is the parastatistics order) at the high energies
of new and future colliders (the Tevatron, the LHC, the NLCs, etc).
\par
Since the paraquantization, carried out at the level of the algebra of
creation and annihilation operators, involves trilinear (or double)
commutation relations in place of the bilinear relations that characterize
Bose and Fermi statistics, sometimes it is highly non-trivial to generalize
some interesting results from ordinary Bose or Fermi statistics to
parastatistics case. In spite of these difficulties, many progresses have been
made in the past years \cite{s4}.
\par
It is also well known that $SU(2)$ and $SU(1,1)$ are the simplest
non-Abelian Lie groups, and the relevant Lie algebras $su(2)$ and $su(1,1)$
have many important applications in various areas of physics. For instance,
the $su(1,1)$ coherent state, has some intrinsic relations to the squeezed
state, one of the most important non-classical states in quantum optics.
Using the ordinary Bose oscillator, one can get a very simple realization of
the $su(1,1)$ algebra. By virtue of the parabose oscillator, one can also do
this, that is, introducing notations $K_{\pm}$ and $K_0$ defined by 
\begin{equation}
K_+ = \frac{1}{2} a^{\dagger 2},~~~K_- = \frac{1}{2} a^2,~~~K_0 = \frac{1}{4}
\left( a^\dagger a + a a^\dagger \right),
\end{equation}
where $a^\dagger$ and $a$ are the parabose creation and annihilation
operators respectively, which satisfy the following trilinear commutation
relations 
\begin{equation}
\left[ a, \{ a^\dagger, a \} \right] = 2a,~~~ \left[ a, a^{\dagger 2}
\right] = 2a^{\dagger},~~~ \left[ a, a^2 \right] =0,
\end{equation}
one has the $su(1,1)$ Lie algebra 
\begin{equation}
\left[ K_0, K_{\pm} \right] = \pm K_{\pm},~~~\left[ K_+, K_- \right] = -2K_0.
\end{equation}
In terms of the generators $K_{\pm}$, the parabose squeezed operator $S(z)$
takes the form of 
\begin{equation}
S(z)=exp \left( \frac{1}{2} za^2 - \frac{1}{2} z^{\ast} a^{\dagger 2}
\right).
\end{equation}

In this letter, using this parabose squeezed operator, we generalize some
interesting results of the ordinary boson systems to systems made of a
single type of parabosons, which include the propagator of a parametric
amplifier in parabose coherent state representation, the explicit form for
parabose squeezed number states and the normalization factors of excitation
states on a parabose squeezed vacuum state.

\section{Propagator of a parabose parametric amplifier}

In this section, we take the parameter $z$ in $S(z)$ as an immaginary number 
$z=-ir$. Using the disentangling theorem of $su(1,1)$ \cite{s5} , we have 
\begin{equation}
S(-ir)= exp \left( -i \,tanh\,r \, \frac{a^{\dagger 2}}{2} \right) exp
\left( ln \,sech\,r \, \frac{\{a^{\dagger},\,a\} }{2} \right) exp \left( -i
\,tanh\,r \, \frac{a^2}{2} \right).
\end{equation}
Differentiating eq.(5) with the parameter $r$ and using the following
operator identities 
\begin{equation}
exp \left( \lambda a^{\dagger 2} \right) \{a^{\dagger},a\} \,exp \left(
-\lambda a^{\dagger 2} \right) = \{a^{\dagger}, a\} - 4 \lambda \,a^{\dagger
2},
\end{equation}
\begin{equation}
exp \left( \lambda \{a^{\dagger}, a\} \right)\, a\, exp \left( -\lambda
\{a^{\dagger}, a\} \right) = a\, e^{-2 \lambda},
\end{equation}
\begin{equation}
exp \left( \lambda a^{\dagger 2} \right)\,a^2\,exp \left( -\lambda
a^{\dagger 2} \right) = a^2 -2 \lambda \{a^{\dagger}, a\} + 4 \lambda^2\,
a^{\dagger 2},
\end{equation}
one gets 
\begin{equation}
\frac{\partial}{\partial r} S(-ir)= -\frac{i}{2} (a^2 + a^{\dagger 2})
S(-ir),
\end{equation}
with a boundary condition $S(r=0)=1$. In fact, eq.(9) also may be derived
from eq.(4) directly by defferentiating $S(-ir)$ with $r$.

Noticing that $S(-ir)$ is a unitary operator, furthermore, when the
parameter $r$ is a function of time $t$ and the initial value of $r$ is
taken as $r(t=t_0)=0$, one may denote $S(-ir)$ as $S(t,t_0)$ and rewrite
eq.(9) as 
\begin{equation}
i \frac{\partial}{\partial t} S(t,t_0) = \frac{1}{2} (a^2 + a^{\dagger 2})
S(t,t_0) \frac{\partial r}{\partial t},
\end{equation}
with the condition $S(t_0,t_0)=1$. Particularly, for $ \partial \,r/
\partial \,t = 2f(t)$, eq.(10) will lead to 
\begin{equation}
i \frac{\partial}{\partial t} S(t,t_0) = f(t) (a^2 + a^{\dagger 2}) S(t,t_0).
\end{equation}
If one introduces a free Hamiltonian $H_0=\frac{\omega}{2} \{a^{\dagger},a\}$
for the parasystem under investigation, and uses eq.(7) and its conjugate
identity, one may rewrite $f(t) (a^2 + a^{\dagger 2})$ as 
\begin{equation}
H_I (t)= f(t)(a^2 +a^{\dagger 2})= exp(iH_0 t) f(t) (a^2 e^{2i\omega t} +
a^{\dagger 2} e^{-2i \omega t} ) exp(-iH_0 t).
\end{equation}
Thus eq.(11) becomes 
\begin{equation}
i \frac{\partial}{\partial t} S(t,t_0) = H_I (t) S(t,t_0),
\end{equation}
which means that one can treat $S(t,t_0)$ and $H_I (t)$ as a
time-displacement operator and the interaction Hamiltonian in the
interaction picture respectively. According to the standard picture
transformation theory in quantum mechanics, a Hamiltonian resulting the
squeezed effect in the Schrodinger picture is 
\begin{equation}
H_S = \frac{\omega}{2} \{a^{\dagger},a\} + f(t) \left( a^2 e^{2i\omega t} +
a^{\dagger 2} e^{-2i\omega t} \right).
\end{equation}
Obviously here $H_S$ is a time-dependent Hamiltonian. When the $a^{\dagger}$
and $a$ in $H_S$ are ordinary bose creation and annihilation operators, $H_S$
describes a kind of non-linear optical phenomena, that is, the interaction
of two types of light beams in a non-linear optical coupler \cite{s6}. $f(t)$
includes a pumping factor (treated as a classical quantity) and another
factor related to two-order of susceptibility of some optical medium.
Especially, when $f(t)=k$, or $r=2k(t-t_0)$, $H_S$ is exactly the
Hamiltonian of a degenerate parametric amplifier. So the $H_S$ in eq.(14) is
a parabose generalization of this kind of Hamiltonian. Also from the
transformation theory between interaction picture and Schrodinger picture,
the time-displacement operator in the Schrodinger picture is 
\begin{equation}
S_S(t,t_0)= e^{-iH_0 t} S(t, t_0) e^{iH_0 t_0},
\end{equation}
which satisfies the following equation 
\begin{equation}
i \frac{\partial}{\partial t} S_S = H_S S_S.
\end{equation}

Now let us consider the propagator of the degenerate parametric amplifier in
a parabose coherent state representation $\langle z|S_S(t,t_0)|z_0 \rangle$,
where $|z \rangle$ is the parabose coherent state given by 
\begin{equation}
|z \rangle = E(|z|^2)^{-1/2} \sum_{n=0}^{\infty} \frac{z^n}{\sqrt{[n]!}} |n
\rangle,
\end{equation}
$|n \rangle$ being the number state of the parabose Fock space 
\begin{equation}
|n \rangle = \frac{a^{\dagger n}}{\sqrt{[n]!}}|0 \rangle,~~~a|n\rangle= 
\sqrt{[n]}|n-1 \rangle,~~~a^{\dagger}|n\rangle=\sqrt{[n+1]}|n+1 \rangle,
\end{equation}
and 
\begin{equation}
[n] = n+ \frac{p-1}{2}(1- (-)^n),~~~E(x)=\sum_{n=0}^{\infty} \frac{x^n}{[n]!}%
,
\end{equation}
where $[n]! =[n][n-1]\cdots [1]$, $[0]!\equiv 1$, and $|0 \rangle$ is the
unique vacuum state of the Fock space, which satisfies $a|0 \rangle=0$ and $%
aa^{\dagger}|0 \rangle = p|0 \rangle$, here $p$ is the parastatistics order (%
$p=1,2,3,\cdots$). Using eq.(7) one has 
\begin{equation}
exp(iH_0 t_0)|z_0 \rangle = e^{ip \omega t_0/2}|z_0e^{i \omega t_0} \rangle.
\end{equation}
Furthermore, according to the disentangling formula (5), it is easily to get
the desired propagator 
\begin{eqnarray}
&&\langle z|S_S(t,t_0)|z_0 \rangle = e^{-ip \omega (t-t_0)/2} 
E(|z|^2)^{-1/2}  E(|z_0|^2)^{-1/2} \nonumber \\
&& \times \left( sech \,2k(t-t_0) \right)^{p/2}
E\left( z^{\ast} z_0 e^{-i \omega (t-t_0)} sech \,
2k(t-t_0) \right)  \nonumber \\
&& \times exp \left( - \frac{i}{2} tanh \,2k(t-t_0) (z_0^2 e^{2i \omega t_0} +
z^{\ast 2} e^{-2i \omega t} ) \right).
\end{eqnarray}
When $p \rightarrow 1$, this propagator will reduce to the ordinary result 
\cite{s6}.

\section{Parabose squeezed number states}

In this section we consider the resulting states from the parabose squeezed
operator $S(r)=exp(\frac{r}{2}a^2-\frac{r}{2}a^{\dagger 2})$ acting on the
parabose number states $|n \rangle$ 
\begin{equation}
|r,n \rangle = S(r) |n \rangle,~~~(n=0,1,2,3, \cdots)
\end{equation}
here for the sake of simplicity, we take the squeezed parameter $z$ as a real
number $r$. We call $|r,n \rangle$ the parabose squeezed number states.
Obviously, $|r,n \rangle$ form a complete and orthonormal state-vector set
for the system made of a single type of parabosons: 
\begin{equation}
\langle r,n|r,m \rangle = \langle n|m \rangle = \delta_{n,m}, ~~~
\sum_{n=0}^{\infty} |r,n \rangle \langle r,n| = 1.
\end{equation}
Using the following transformations 
\begin{eqnarray}
S(r)\,a\,S(r)^{-1} &=& cosh \,r\,a + sinh \,r\,a^{\dagger},  \nonumber \\
S(r)\,a^{\dagger}\,S(r)^{-1} &=& cosh \,r\,a^{\dagger} + sinh \,r\,a
\end{eqnarray}
and the disentangling formula (5), we have 
\begin{equation}
|r,n \rangle = \frac{(sech \,r)^{p/2}}{\sqrt{[n]!}} \left(
cosh \,r\,a^{\dagger}
+ sinh \,r\,a \right)^n e^{-tanh \,r\,a^{\dagger 2}/2} |0 \rangle.
\end{equation}
In terms of the deformed Hermite polynomials $H_n^{(p)} (x)$ (see the
appendix), the explicit form of $|r,n \rangle$ is 
\begin{equation}
|r,n \rangle = \frac{(sech \,r)^{p/2}}{\sqrt{[n]!}} \left(-\frac{tanh \,r}{2}
\right)^{n/2} H_n^{(p)} \left( \frac{a^{\dagger}}{i\sqrt{sinh \,2r}} \right)
e^{-tanh \,r\,a^{\dagger 2}/2} |0 \rangle.
\end{equation}

We prove eq.(26) by induction. Firstly, from (25), and using 
\begin{equation}
e^{tanh \,r\,a^{\dagger 2}/2}\,a^n\,e^{-tanh \,r\,a^{\dagger 2}/2} = (a -
tanh \,r\,a^{\dagger} )^n,
\end{equation}
for $n=1$ case, we have 
\begin{eqnarray}
|r,1 \rangle &=& \frac{(sech \,r)^{p/2}}{\sqrt{[1]!}\,cosh \,r} a^{\dagger}
e^{-tanh \,r\,a^{\dagger 2}/2} |0 \rangle  \nonumber \\
&=& \frac{(sech \,r)^{p/2}}{\sqrt{[1]!}} \left( -\frac{tanh \,r}{2}
\right)^{1/2} H_1^{(p)} (\chi) e^{-tanh \,r\,a^{\dagger 2}/2} |0 \rangle,
\end{eqnarray}
where $\chi$ stands for $a^{\dagger}/i\sqrt{sinh \,2r}$. Then supposing 
\begin{equation}
|r,n-1 \rangle = \frac{(sech \,r)^{p/2}}{\sqrt{[n-1]!}} \left(
-\frac{tanh \,r}{2} \right)^{\frac{n-1}{2}}H_{n-1}^{(p)} (\chi)
e^{ -tanh \,r\,a^{\dagger 2}/2} |0 \rangle,
\end{equation}
we will have 
\begin{eqnarray}
&& |r,n \rangle =\frac{1}{\sqrt{[n]}} \left( cosh \,r\,a^{\dagger}+
sinh \,r\,a \right) |r,n-1 \rangle  \nonumber \\
&& = \frac{(sech \,r)^{p/2}}{\sqrt{[n]!}} \left( -\frac{tanh \,r}{2}
\right)^{\frac{n-1}{2}} cosh \,r\,a^{\dagger}\,H_{n-1}^{(p)} (\chi)
e^{-tanh \,r\,a^{\dagger 2}/2} |0 \rangle  \nonumber \\
&&\,\, + \frac{(sech \,r)^{p/2}}{\sqrt{[n]!}} \left( -\frac{tanh \,r}{2}
\right)^{ \frac{n-1}{2}} sinh \,r\,e^{-tanh \,r\,a^{\dagger 2}/2}
(a- tanh \,r\,a^{\dagger}) \nonumber \\
&&\,\,\times H_{n-1}^{(p)}(\chi) |0 \rangle.
\end{eqnarray}
By use of the relations 
\begin{eqnarray}
\left[a,\,a^{\dagger n}\right]&=& a^{\dagger n-1}\left(n+\frac{p-1}{2}
(1-(-)^n)R \right), \nonumber\\
\left[a^{\dagger},\,a^n\right]&=&-a^{n-1}\left(n+\frac{p-1}{2}(1-(-)^n)R
\right),
\end{eqnarray}
where $R$ is a reflection operator (see eq.(38)), it is easily to find that 
\begin{equation}
a\,H_{n-1}^{(p)} (\chi) |0 \rangle = \frac{2[n-1]}{i\sqrt{sinh \,2r}}
H_{n-2}^{(p)} (\chi) |0 \rangle.
\end{equation}
Substituting eq.(32) into (30), we obtain 
\begin{eqnarray}
|r,n \rangle &=& \frac{(sech \,r)^{p/2}}{\sqrt{[n]!}} \left(
-\frac{tanh \,r}{2} \right)^{n/2} \left( 2\chi H_{n-1}^{(p)}(\chi) -
2[n-1] H_{n-2}^{(p)} (\chi) \right)  \nonumber \\
& & \times e^{-tanh\,r\,a^{\dagger 2}/2} |0 \rangle.
\end{eqnarray}
At last, by virtue of a recursion relation for the deformed Hermite
polynomials (see the appendix) 
\begin{equation}
H_{n+1}^{(p)} (x) - 2x H_n^{(p)} (x) + 2[n] H_{n-1}^{(p)} (x) = 0,
\end{equation}
eq.(26) is proved.

\section{Excitations on a parabose squeezed vacuum state}
In this section we consider the excitations on a parabose squeezed vacuum
state which defined by 
\begin{equation}
||n,r \rangle = a^{\dagger n} |0,r \rangle,
\end{equation}
where $|0,r \rangle = S(r) |0 \rangle$ is the parabose squeezed vacuum
state. It is easily to see that $|0,r \rangle$ is normalized, however, the
states $||n,r \rangle$ have not been normalized. We would like to point out
that the states $||n,r \rangle$ may be normalized in terms of the deformed
Legendre polynomials $P_n^{(p)}(x)$ (see the appendix) as 
\begin{equation}
\langle n,r||n,r \rangle = [n]! (cosh \,r)^n P_n^{(p)} (cosh \,r).
\end{equation}
We prove eq.(36) by induction again. Firstly, we have 
\begin{eqnarray}
a |0,r \rangle &=& (sech \,r)^{p/2} a\,e^{-tanh \,r \,a^{\dagger 2}/2}
|0 \rangle \nonumber \\
&=& (sech \,r)^{p/2} e^{-tanh \,r\,a^{\dagger 2}/2} (a -tanh \,r\,
a^{\dagger})|0 \rangle  \nonumber \\
&=& -tanh \,r\,a^{\dagger} |0,r \rangle .
\end{eqnarray}
By virtue of the so-called $R$-deformed Heisenberg algebra \cite{s7}, one
can rewrite the trilinear commutation relations (2) as 
\begin{equation}
[a,\,a^{\dagger}]=1+(p-1)R,~~~\{R,\,a\}=\{R,\,a^{\dagger}\}=0,~~~R^2=1,
\end{equation}
where $R$ is the reflection operator. Using the $R$-deformed commutation
relation (38) and noticing that $R |0 \rangle = |0 \rangle$, we find 
\[
\langle 1,r||1,r \rangle = \langle 0,r|a a^{\dagger}|0,r \rangle = \langle
0,r| \left( 1 + a^{\dagger} a + (p-1)R \right) |0,r \rangle 
\]
\begin{equation}
= p + tanh^{2}{r} \langle 1,r||1,r \rangle = [1]!cosh \,rP_1^{(p)} (cosh \,r).
\end{equation}
Then supposing eq.(36) is true for $m \leq n$, that is, 
\begin{equation}
\langle m-1,r||m-1,r \rangle = [m-1]!(cosh \,r)^{m-1} P_{m-1}^{(p)}(cosh \,r),
\end{equation}
we show that eq.(36) works. In fact, using eq.(31) we have 
\begin{eqnarray}
& &\langle n,r||n,r \rangle = \langle 0,r|a^{n-1}aa^{\dagger} a^{\dagger
(n-1)} |0,r \rangle  \nonumber \\
& &= \langle 0,r|a^{n-1} \left( 1+a^\dagger a + (p-1)R \right) a^{\dagger
(n-1)} |0,r \rangle  \nonumber \\
& &= \langle n-1,r||n-1,r \rangle +(p-1)(-)^{n-1} \langle n-1,r||n-1,r \rangle
\nonumber \\
& &\,\,+ \langle 0,r| \left( a^{\dagger} a^{n-1} + [n-1]a^{n-2} \right)
\left( a^{\dagger (n-1)} a + [n-1]a^{\dagger (n-2)} \right) |0,r \rangle 
\nonumber \\
& &=tanh^{2}{r} \langle n,r||n,r \rangle -[n-1]^2 \langle n-2,r||n-2,r \rangle
\nonumber\\
& &\,\, + \left( 2[n-1] +1 +(p-1)(-)^{n-1}
\right)  \langle n-1,r||n-1,r \rangle,
\end{eqnarray}
or 
\begin{eqnarray}
& &\langle n,r||n,r \rangle = - cosh^{2}{r} [n-1]^2 \langle n-2,r||n-2,r
\rangle  \nonumber \\
& & \,\,+ cosh^{2}{r} \left( 2[n-1] +1 + (p-1)(-)^{n-1} \right) \langle
n-1,r||n-1,r \rangle .
\end{eqnarray}
Substituting (40) into (42), we get 
\begin{eqnarray}
& &\langle n,r||n,r \rangle = - cosh^{n}{r} [n-1]![n-1]P_{n-2}^{(p)}(cosh \,r)
\nonumber \\
& &\,\,+cosh^{n+1}{r}[n-1]!\left( 2[n-1] +1 +(p-1)(-)^{n-1} \right)
P_{n-1}^{(p)} (cosh \,r) .
\end{eqnarray}
Noticing that $2[n-1] +1 + (p-1)(-)^{n-1} =[2n-1]$ and using the following
recursion relation for $P_n^{(p)} (x)$ (see the appendix), 
\begin{equation}
[n+1] P_{n+1}^{(p)} (x)-[2n+1]xP_n^{(p)} (x)+[n]P_{n-1}^{(p)} (x) =0,
\end{equation}
we finally arrive at eq.(36). Thus we see that the deformed Legendre
polynomials indeed can be used to normalize the excitation states on a
squeezed vacuum state for a single parabose mode.

In summary, in this letter, we generalize some interesting results from the
ordinary Bose statistics to the parabose statistics which are related to the
squeezed operator. In these generalizations, some deformed polynomials and
the relevant recursion relations are used which will be defined and explained
in the appendix.

\section*{Appendix}
\subsection*{Deformations of ordinary Hermite and Legendre polynomials}
In order to explain the definitions and properties of the new kind of
deformation of the ordinary Hermire polynomials and Legendre polynomials,
let us introduce a kind of deformation of the ordinary derivative
operator, which was proposed for developing coordinate representation
theory of parabose system \cite{s8}. The deformed derivative operator $D$ is
defined by 
\begin{eqnarray}
D f(x) &\equiv& \frac{D}{Dx} f(x) = \frac{d}{dx} f(x) + \frac{p-1}{2x} (1-R)
f(x)  \nonumber \\
&=& d\,f(x) + \frac{p-1}{2x} (f(x)-f(-x)),
\end{eqnarray}
where $df=df/dx$ and $R$ is the reflection operator which has a property
$R\,f(x)=f(-x)$ for any $x$ dependent function $f(x)$. Let us consider
solutions of a second-order differential equation based on the deformed
derivative operator $D$
\begin{equation}
D^2 f(x) -2xDf(x)+ \mu f(x)=0.
\end{equation}
When the parameter $\mu$ takes eigenvalues $\mu =2[n], n= 0,1,2,3, \cdots$,
for each given parastatistics order $p$, eq.(46) has the following solutions 
\begin{equation}
H_n^{(p)} (x)=[n]!\sum_{k=0}^{[n/2]^{^{\prime}}} \frac{(-)^k (2x)^{n-2k}}{k!
[n-2k]!},
\end{equation}
where $[k]^{^{\prime}}$ in the above of $\sum$ stands for the largest
integer smaller than or equal to $k$. We call polynomials (47) the deformed
Hermite polynomials because when $p \rightarrow 1$ (47) will reduce to the
ordinary Hermite polynomials. The first few of $H_n^{(p)}(x)$ have the
following explicit forms
\begin{eqnarray}
& &H_0^{(p)}(x)=1,~~~H_1^{(p)}(x)=2x,~~~H_2^{(p)}(x)=4x^2-[2]! \nonumber\\
& &H_3^{(p)}(x)=8x^3-4[3]x,~~~\cdots
\end{eqnarray}
In terms of the deformed operator $D$, $H_n^{(p)}(x)$ also have its
differential form
\begin{equation}
H_n^{(p)} (x)=(-)^ne^{x^2} D^n e^{-x^2}.
\end{equation}
The generating function of $H_n^{(p)} (x)$ is 
\begin{equation}
e^{-t^2} E(2tx)=\sum_{n=0}^{\infty} \frac{t^n}{[n]!} H_n^{(p)} (x).
\end{equation}
There are some definite relations between neighbouring $H_n^{(p)} (x)$ and
their derivatives. The main recursion relations are the following two: 
\begin{equation}
DH_n^{(p)}(x) -2[n]H_{n-1}^{(p)}(x)=0,
\end{equation}
\begin{equation}
H_{n+1}^{(p)}(x)-2xH_n^{(p)}(x)+2[n]H_{n-1}^{(p)} (x)=0.
\end{equation}
\par
Another second-order differential equation also based on the deformed
operator $D$ 
\begin{equation}
(1-x^2)D^2 f(x)-2xD f(x)+\mu f(x)=0
\end{equation}
will lead to the notion of the deformed Legendre polynomials when $%
\mu=[n][n+1], n=0,1,2,3, \cdots$, 
\begin{equation}
P_n^{(p)}(x)= \sum_{k=0}^{[n/2]^{^{\prime}}} \frac{(-)^k [2n-2k]! x^{n-2k}}{%
2^n k! (n-k)! [n-2k]!}.
\end{equation}
The first few of $P_n^{(p)}(x)$ are
\begin{eqnarray}
& &P_0^{(p)}(x)=1,~~~P_1^{(p)}(x)=x,~~~P_2^{(p)}(x)=\frac{1}{2}([3]x^2-[1]),
\nonumber\\
& &P_3^{(p)}(x)=\frac{1}{2}([5]x^3-[3]x),~~~\cdots.
\end{eqnarray}
A defferential expression for $P_n^{(p)}(x)$ is
\begin{equation}
P_n^{(p)}(x)= \frac{1}{2^n n!} D^n (x^2-1)^n.
\end{equation}
The main recursion relations of $P_n^{(p)}(x)$ are
\begin{equation}
[n+1]P_{n+1}^{(p)} (x)-[2n+1]x P_n^{(p)}(x)+[n]P_{n-1}^{(p)}(x)=0,
\end{equation}
\begin{equation}
DP_{n+1}^{(p)}(x)-xDP_n^{(p)}(x)-[n+1]P_n^{(p)}(x)=0,
\end{equation}
\begin{equation}
xDP_n^{(p)}(x)-DP_{n-1}^{(p)}(x)-[n]P_n^{(p)}(x)=0,
\end{equation}
\begin{equation}
DP_{n+1}^{(p)}(x)-DP_{n-1}^{(p)}(x)-[2n+1]P_n^{(p)}(x)=0,
\end{equation}
\begin{equation}
(x^2 -1)DP_n^{(p)}(x)-[n]xP_n^{(p)}(x)+[n]P_{n-1}^{(p)}(x)=0.
\end{equation}

\bigskip
{\sl Note added in proof.} The authors would like to thank one of the referees
for drawing to their attention a paper by Saxena and Mehta \cite{s9}, in which
the parabose squeezed vacuum state wes discussed, and in the present work the
excitations on the parabose squeezed vacuum state was considered and
normalized in terms of a new kind of deformed Legendre polynomials.


\bigskip

\begin{thebibliography}{9}
\bibitem{s1}  H. S. Green, {\em Phys.Rev.} {\bf 90} (1953) 270.

\bibitem{s2}  B. I. Halperin, {\em Phys.Rev.Lett.} {\bf 52} (1984) 1583.

\bibitem{s3}  F. Wilczek, {\em Fractional Statistics and Anyon 
Superconductivity} (World Scientific, Singapore, 1990).

\bibitem{s4}  Y. Ohnuki and S. Kamefuchi, {\em Quantum Field Theory and 
Parastatistics} (Springer-Verlag, 1982).

\bibitem{s5}  D. R. Truax, {\em Phys.Rev.} D{\bf 31} (1985) 1988.

\bibitem{s6}  H. Fan, {\em Representation and Transformation Theory in Quantum
Mechanics} (Shanghai Scientific and Technical Publishers,1997).

\bibitem{s7}  M. A. Vasiliev, {\em Int.J.Mod.Phys.} {\bf A6} (1991) 1115; 
T. Brzezinski, I. L. Egusquiza and A. J. Macfarlane, {\em Phys.  Lett.} {\bf %
B 311} (1993) 202;  M. S. Plyushchay, {\em Nucl.Phys.} {\bf B491} (1997) 619.

\bibitem{s8}  S. Jing, {\em J.Phys.A:Math.Gen.} {\bf 31} (1998) 6347.

\bibitem{s9}  G. M. Saxena and C. L. Mehta, {\em J.Math.Phys.} {\bf 32} (1991)
783.

\end{thebibliography}
\end{document}